\documentclass[useAMS,usenatbib, usegraphicx]{mn2e}
\usepackage{color}
\usepackage{amssymb}

\title[Weak shock-cloud interaction in 3D]
{Shock-triggered formation of magnetically-dominated clouds.\\ 
II. Weak shock-cloud interaction in three dimensions.}
\author[S. Van Loo et al.]
{S. Van Loo$^{1,2}$,\thanks{E-mail: svenvl@astro.ufl.edu} 
S. A. E. G. Falle$^3$ and T. W. Hartquist$^1$ \\
$^1$ School of Physics and Astronomy, University of Leeds, Leeds LS2 9JT, UK\\
$^2$ Department of Astronomy, University of Florida, Gainesville, Florida 
32611, USA\\
$^3$ Department of Applied Mathematics, University of Leeds, Leeds LS2 9JT, UK}
\begin{document}

\date{Accepted - . Received - ; in original form -}

\pagerange{\pageref{firstpage}--\pageref{lastpage}} \pubyear{2006}

\maketitle

\label{firstpage}

\begin{abstract}
To understand the formation of a magnetically dominated molecular
cloud from an atomic cloud, we study the interaction of a weak, radiative
shock with a magnetised cloud. The thermally stable warm atomic cloud 
is initially in static equilibrium with the surrounding hot ionised gas. 
A shock propagating through the hot medium then interacts with the cloud.
We follow the dynamical evolution of the shocked cloud with a time-dependent
ideal magnetohydrodynamic code. By performing the simulations in 3D, we
investigate the effect of different magnetic field orientations including 
parallel, perpendicular and oblique to the shock normal. We find that 
the angle between the shock normal and the magnetic field must be small 
to produce clouds with properties similar to observed molecular clouds. 
\end{abstract}


\section{Introduction}
Molecular clouds exhibit a hierarchical density structure 
\citep[e.g.][]{BS86}. Stars form in the densest regions, or dense cores, 
which become gravitationally unstable. In fact, molecular clouds in the Solar 
neighbourhood that don't harbour any stars are rare. While most of the 
stars within molecular clouds are young ($\approx $ 1-2 Myr), stellar 
associations older than 5 Myr are devoid of molecular gas \citep[e.g.][]{BH07}. 
This suggests that molecular clouds are short-lived, transient objects
and that the time-lag between cloud formation and stellar birth is short.
 
The rapid onset of star formation requires that large density contrasts
arise while the parental cloud forms. It has been suggested that this 
fragmentation results from thermal processes in the interstellar medium 
(ISM) \citep[see e.g.][and references therein]{KKH09}. For a range of
pressures and the heating and cooling rates appropriate for diffuse atomic 
gas, two thermally stable phases exist, i.e. a rarefied, warm phase and 
a cold, dense phase, which can co-exist in pressure equilibrium 
\citep{FGH69,Wetal95}. Atomic gas at intermediate temperatures is 
subject to a thermal instability. A sufficient rise in pressure causes the 
cold phase to be the only stable one \citep{F65}. 

Previous studies of molecular cloud formation due to thermal
instability mainly focus on collisions of warm gas streams in the context of 
expanding supernovae shells or spiral arm shocks 
\citep[e.g.][]{Hetal08, HSH09, II09}. Beside being thermally unstable in 
some circumstances, the collision region is prone to numerous dynamical 
instabilities such as the Kelvin-Helmholtz, Rayleigh-Taylor and nonlinear thin 
shell or Vishniac instabilities. In the turbulent shocked layer 
cold gas clumps then arise on short timescales. Although the derived density 
and velocity structures depend strongly on the magnitude and orientation of the 
magnetic field, they resemble the ones observed in molecular clouds and diffuse 
HI clouds \citep{HSH09}. 

While this model of flow-driven structure generation shows rapid onset of 
star formation while the parental cloud is forming, it cannot account for 
the observed low star formation rates \citep{KKH09}. As there is a 
continuous instream of gas into 
the collision region, too much of the accumulated gas will be converted into 
stars. For the same reason, it also cannot explain cloud lifetimes. 
Some of these limitations do not necessarily occur when studying the same 
processes in cloud-cloud interactions \citep{KW98,Metal99} or shock-triggered 
models where shocks overrun warm, diffuse density perturbations 
\citep[][hereafter Paper I]{IK04,PaperI}. Indeed, numerical models of shocks 
interacting with clouds show that the clouds fragment due to dynamical 
instabilities \citep{Metal94}, thereby setting a limit on both the cloud 
lifetime and the star formation rate. 

Colliding flow-driven models only form thin filamentary clouds, while 
different cloud morphologies such as a cometary cloud structure
with a massive head and long-spread tail, are also observed \citep{Tetal02}. 
Such a morphology is characteristic for clouds harbouring cluster-forming cores.
The W3 Giant Molecular Cloud (GMC) is an example of such a cloud 
\citep{Metal07}.  As there is no constraint on the geometry of 
density perturbations in the shock-triggered models several morphologies 
can be reproduced. 

The interaction of a shock with a magnetised cloud has been studied 
extensively in a two-dimensional (2D) axisymmetric geometry, both for adiabatic 
\citep[e.g.][]{Metal94,Netal06} and radiative shocks 
\citep[e.g.][Paper I]{Fetal05}. However, these simulations 
are limited to a single configuration of the magnetic field, i.e. 
the magnetic field is parallel to the symmetry axis and the shock normal. 
To study the effect of the magnetic field orientation, it is necessary to 
model the shock-cloud interaction in 3D. Although such simulations
have been around for some time \citep{SN92,Getal00}, only recently high enough 
resolution has been achieved for general cloud properties to converge 
\citep{SSS08}. However, the Shin et al. study focuses on strong, adiabatic 
shocks, while  the results of Paper I show that molecular clouds most likely 
form from interactions involving weak, radiative shocks. The first 3D 
simulations of the interaction of a radiative shock with a magnetised cloud 
were performed by \citet{Letal09}, although they use a nearly isothermal 
equation of state to simulate strong radiative cooling. 

In this paper we study the interaction of a weak, radiative shock with a 
magnetised cloud. We focus on the early stages of the evolution before the 
cloud re-expands and fragments. In Sect.~\ref{sect:model}  we describe the 
numerical code and the initial conditions. The results are presented 
in Sect.~\ref{sect:results}, while we discuss the cloud properties in 
\ref{sect:properties}. Finally, we finish the paper with a summary 
(Sect.~\ref{sect:conclusions}).

\section{Numerical model}\label{sect:model}
\subsection{Numerical code} 
To solve the ideal magnetohydrodynamics equations we use a second-order 
Godunov scheme with a linear Riemann solver \citep{F91}. To ensure that 
the solenoidal constraint is met, a divergence cleaning algorithm is 
implemented in the numerical scheme \citep{Detal02}. The thermal behaviour 
of the diffuse atomic gas as described by \citet{Wetal95} is treated 
through the inclusion of a source term in the energy equation. 
The exact expressions for the 
cooling and heating function are given in Paper I and are from \citet{Setal02}.
As the cooling time scale can be much smaller
than the dynamical time scale, the equations are stiff. By using an 
exponential time differencing method the numerical scheme remains 
stable for the larger time step set by the Courant condition 
\citep[see e.g.][]{T06}.

\subsection{Initial conditions}
In our models, a quiescent, uniform and spherical cloud of radius 
(R$_{cl}$) 200 pc and number density $n = 0.45$ cm$^{-3}$ is initially in 
thermal equilibrium. The gas is in the thermally stable warm phase with 
a thermal pressure $p = 2825k$ where $k$ is the Boltzmann constant. 
Also, the cloud is in pressure equilibrium 
with the surrounding hot ionised medium (n = 0.01 cm$^{-3}$).
The analysis of \citet{BM90} shows that the hot component
of the interstellar medium (ISM) is thermally stable due to the subsequent 
reheating by supernovae. Thus, we assume that the surrounding gas cools 
adiabatically. 

While observations show that molecular clouds are magnetically dominated with 
values of the ratio $\beta$ of the thermal gas pressure to magnetic pressure
of the order 0.04-0.6, most diffuse clouds have weak magnetic fields 
\citep{CHT03}. On scales large compared to the size of clouds, 
the magnetic and thermal gas pressures are 
actually comparable. Therefore, we adopt a uniform magnetic field with a 
strength such that $\beta = 1$. (For our models this means that the 
magnetic field strength is roughly 1~$\mu$G.)

As in Paper I, we consider a steady, planar shock hitting the quiescent 
cloud.  The shock is propagating in the negative $z$-direction with the 
angle between the shock normal and the magnetic field either 0$^o$ 
(parallel model), 15$^o$, 45$^o$ (oblique models) or 90$^o$ (perpendicular 
model). The magnetic field is in the $x-z$ plane for the oblique models and 
in the $x$-direction for the perpendicular model. The shock velocity 
through the hot ionised medium $v_{ext}$ is 2.5 times the hot gas sound 
speed $a$ (i.e. the shock sonic Mach number is 2.5). As the fast 
magnetosonic speed $c_f$ lies within the range of values 
$a < c_f < \sqrt{2}a$, this shock is a weak, fast-mode shock.

\subsection{Computational domain}
The computational domain is $-2 R_{cl} < x,y < 2R_{cl} $, 
and $-2.066 R_{cl} < z < 3 R_{cl}$ with free-flow boundary conditions on 
all boundaries except on the positive $z$ boundary. There we fix the 
variables to have the postshock flow conditions calculated from the 
adiabatic Rankine-Hugionot relations.

For models of adiabatic shock-cloud interaction, it is necessary to have at 
least $\approx 100$ grid points per cloud radius $R_{cl}$ to obtain convergence 
of general cloud properties like the shape of the cloud and the rms 
velocities along each axis \citep{Metal94,SSS08}. However, some cloud 
properties do change with increasing resolution. These are usually associated 
with quantities that are sensitive to small-scale processes. 
For radiative shocks, \cite{YFC09} show that there is no true convergence 
at 100 cells per cloud radius. Changes induced by an increasing resolution
have a global effect later on in the dynamical evolution of the cloud.
It is likely that convergence is only achieved when the cooling lengths 
are adequately resolved. At the moment, however, it is not feasible to
perform simulations of a shock-cloud interaction in 3D that attain the
required resolution. Therefore, we adopt a resolution similar to that used 
by \cite{SSS08} for the adiabatic shocks. We use 
an adaptive mesh with 5 levels of refinement with the resolution at 
the finest grid 480$^2 \times$ 608. Such a resolution corresponds to a physical grid 
spacing of 1.58 pc. As translucent clumps in GMCs have  length scales of
about 3~pc, it is clear that we cannot resolve such structures. Our 
simulations, thus, can only show the onset of clump formation.

As the cloud is accelerated to move with the post-shock flow 
\citep[e.g.][]{Metal94}, the cloud eventually moves off the grid. To avoid this
we calculate the density-weighed average velocity of the cloud along each
different axis at every timestep. The density-weighed average of 
each velocity component is defined as 
\begin{equation}\label{eq:average}
    <f> = \frac{1}{M_{cl}}\int_{V}\rho C f dV,
\end{equation}
where $f$ is $v_x$, $v_y$ or $v_z$ and $C$ is a scalar which is 1 for cloud 
material and 0 for ambient gas. $\rho, V$ and $M_{cl}$ are the mass density,
volume, and cloud mass. By performing a Galilean transformation 
of the computational domain, the cloud remains in the centre of the grid. 
Furthermore, we are able to study the acceleration of the cloud. Note that  
density-weighed averages can also be calculated for other flow variables. 

\begin{figure*}
\includegraphics[width = 8.4 cm]{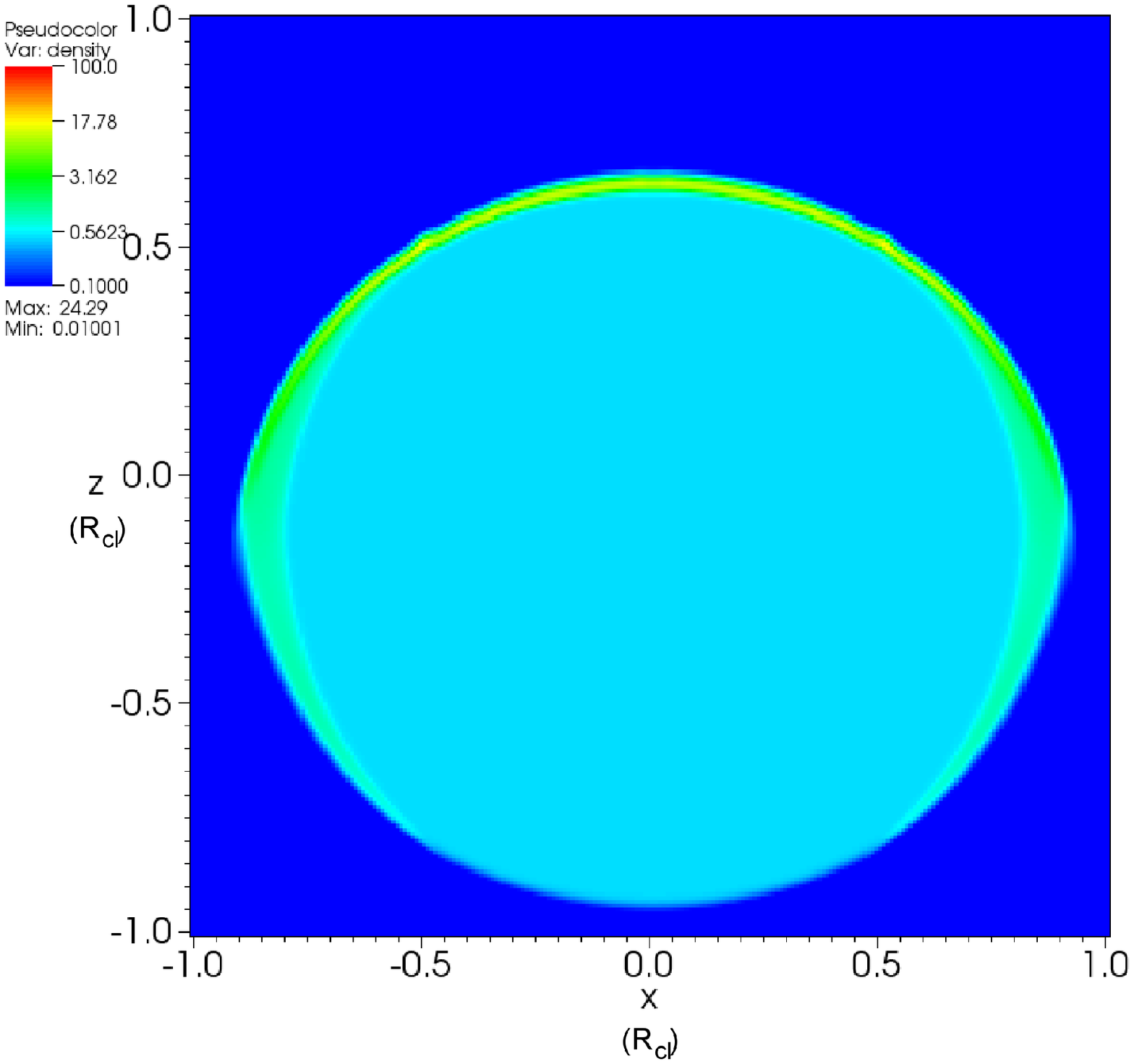}
\includegraphics[width = 8.4 cm]{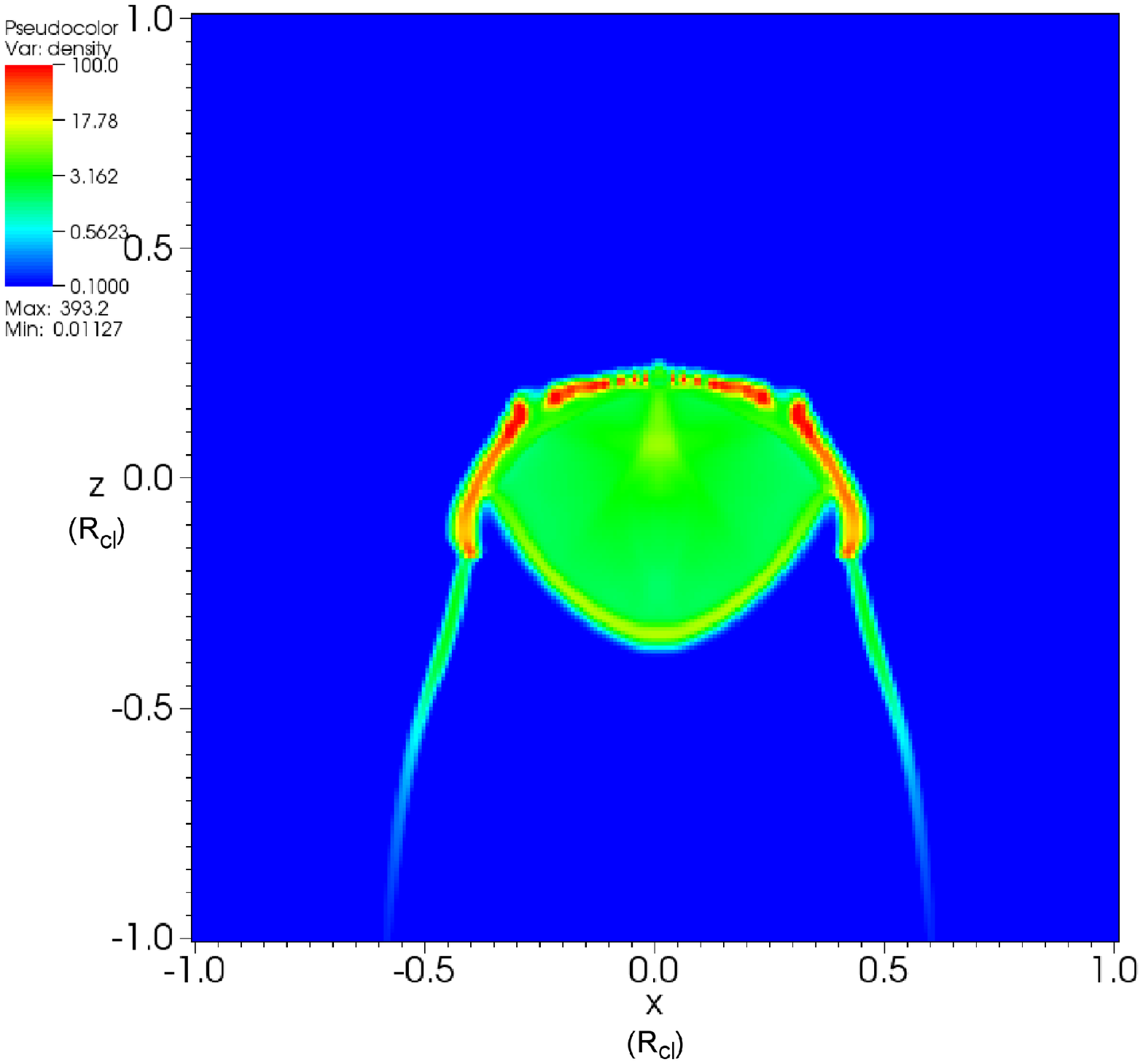}
\caption{
Slice of the logarithmic number density in the $y=0$ plane for the  
parallel shock model. The left panel shows the density distribution 
at 2.5 Myr (or just after t$_{cp}$) and
the right panel at 9 Myr (or just after t$_{cc}$).
The value of the number density is between 0.1 and 100 and the 
units of the axes are in $R_{cl}$. 
}
\label{fig:paboundary}
\end{figure*}

\section{Dynamical evolution}\label{sect:results}
\subsection{Parallel shock}\label{sect:parallel}
The shock-cloud interaction for a parallel shock in 3D
is nearly identical to the evolution studied in the 2D axisymmetric 
simulations of Paper~I. It is useful to describe the dynamical 
evolution here again, not only to show the differences compared to the 2D
results, but also because some of the dynamical characteristics are 
relevant for the perpendicular and oblique models.

\begin{figure*}
\includegraphics[width = 7.8 cm]{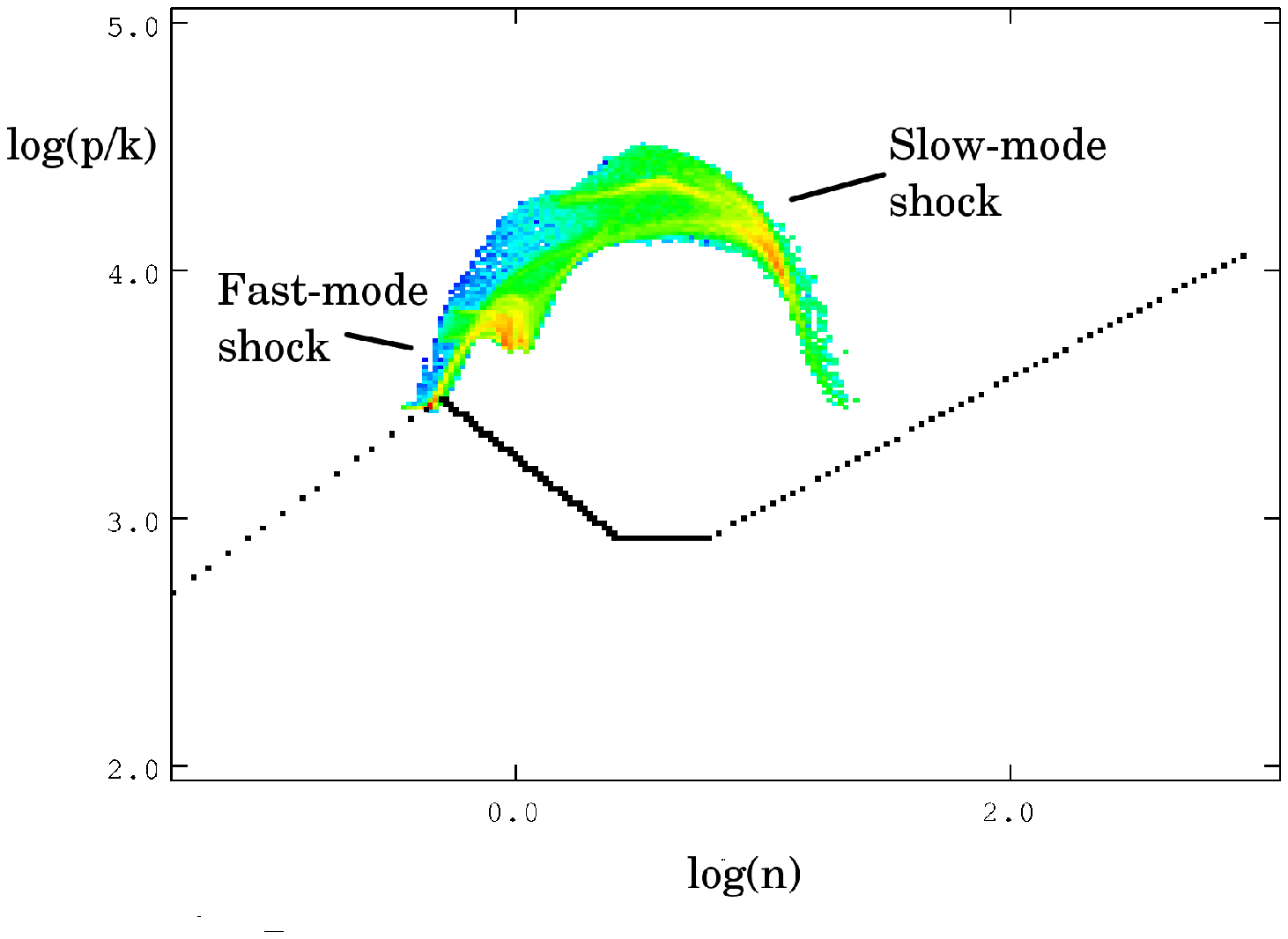}
\includegraphics[width = 7.8 cm]{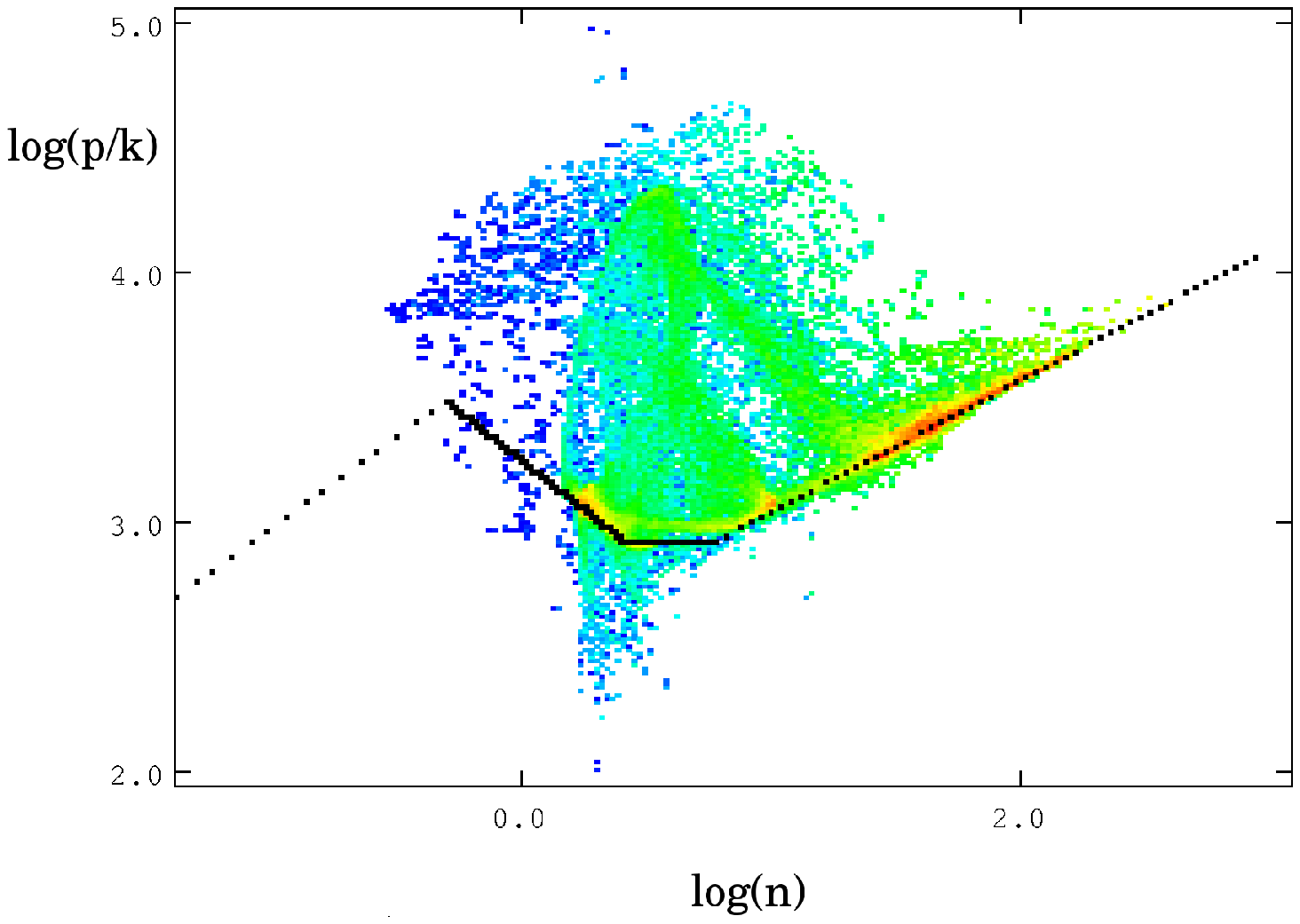}
\includegraphics[width = 1.4 cm]{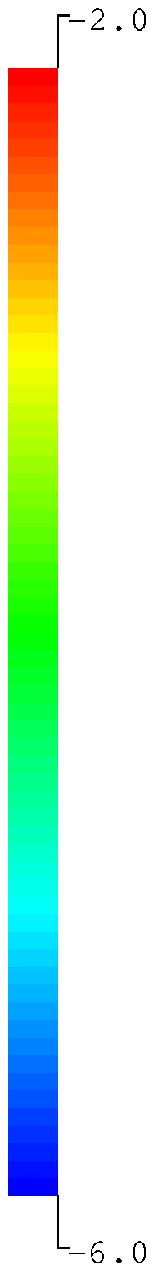}
\caption{Distribution of the mass fraction in a phase diagram for the 
parallel shock model. The left panel shows the phase diagram at 2.5 Myr
(or just after t$_{cp}$) and the right panel at 9 Myr (or just after
t$_{cc}$). The values for the mass fraction range between 10$^{-6}$ and 
10$^{-2}$. The black line represents the equilibrium curve.}
\label{fig:pa_phase}
\end{figure*}

Figure~\ref{fig:paboundary} shows results for the parallel case at two 
times.  As the intercloud shock sweeps around the cloud (in a time of 
$t_{cp} = 2R_{cl}/v_{ext}$), a shock is transmitted into the cloud and 
a bow-shock forms in front of the cloud. The transmitted fast-mode shock 
has a lower propagation speed than the intercloud shock, i.e. 
$v_{int} = v_{ext}/\chi^{1/2}$ where $\chi$ is the density ratio of 
cloud/intercloud gas. Therefore, a velocity shear layer forms  
at the cloud boundary. Because of this slip surface, a vortex ring develops
and sweeps cloud material away from the cloud (see Fig.~\ref{fig:paboundary}). 

\begin{figure*}
\includegraphics[width = 7.8 cm]{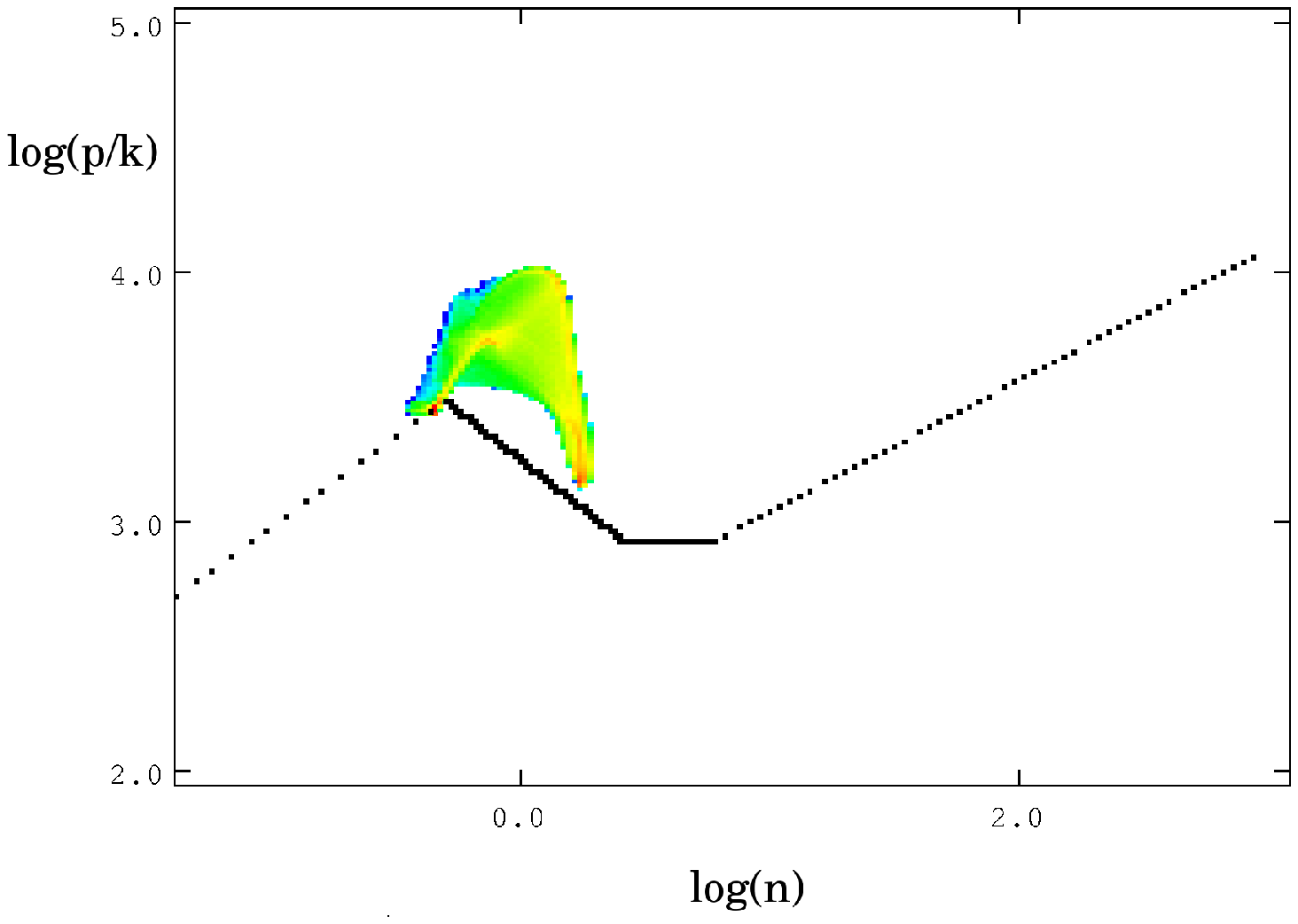}
\includegraphics[width = 7.8 cm]{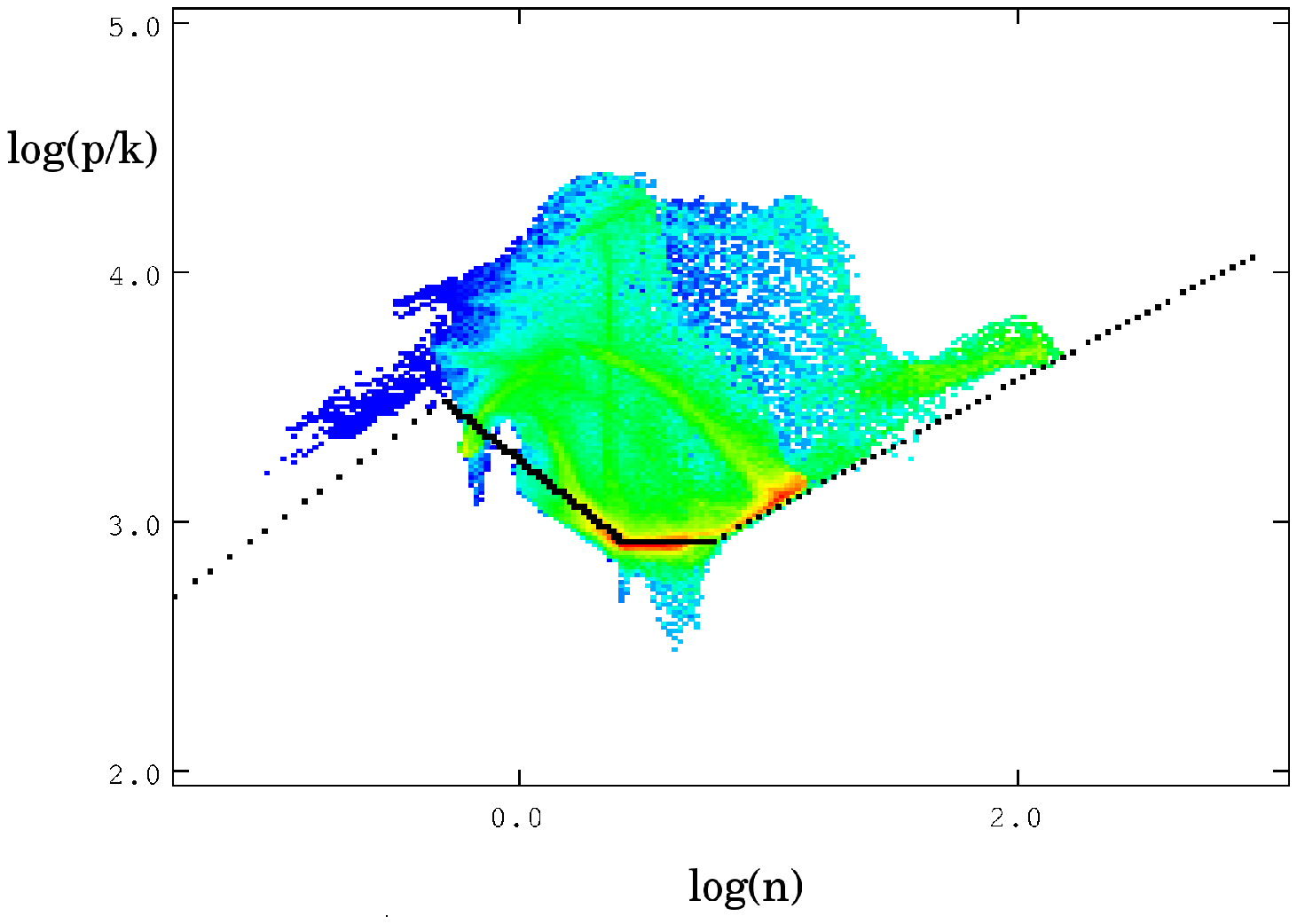}
\includegraphics[width = 1.4 cm]{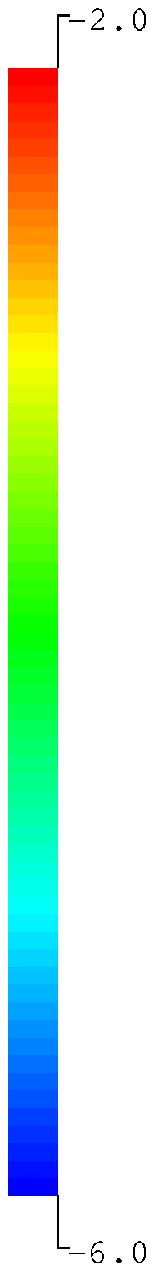}
\caption{
Similar to Fig.~\ref{fig:pa_phase} but for the perpendicular shock model.}
\label{fig:pe_phase}
\end{figure*}

As the fast-mode shock moves through the cloud, it compresses and heats 
the initially thermally-stable warm gas so that it ends up being thermally 
unstable. Figure.~\ref{fig:pa_phase} shows the distribution of mass in the 
$p-n$ phase space at two different times. The time scale for 
radiative cooling is shorter than the propagation time of the fast-mode 
shock through the cloud, given by the cloud-crushing time 
$t_{cc} = R_{cl}/v_{int}$. Thus, the gas loses a significant fraction of 
its internal energy during the compression. Furthermore, the magnetic 
pressure increases behind the fast-mode shock.  Hence, the value of $\beta$ 
drops below unity inside the cloud. This provides the ideal conditions
for the generation, by MHD waves, of dense clumps and cores
\citep{FH02, VFH06, VFH08}. Cold, dense clumps can also form due to small 
perturbations along the unstable part of the equilibrium curve \citep{KI06}. 
In our simulation we find a small fraction of cloud material on the 
thermally unstable part of the equilibrium curve after $\approx$ 6 Myr 
(or 0.7t$_{cc}$). Typical timescales for both processes are a few Myr. 
As the cloud flattens and fragments in about 1.5-2~t$_{cc}$ (see Paper I), 
there is ample time for these processes to work. However, the numerical 
resolution of our simulations is insufficient to establish whether these 
processes are the dominant formation mechanisms of dense clumps and 
cores within clouds.
As we cannot follow this formation process, we stop the simulation 
shortly after $t_{cc}$. This timescale was chosen because the re-expansion
phase of the cloud then starts. Also, self-gravity which is not included 
in these simulations becomes globally important around this time and 
will affect the subsequent dynamical evolution. 

The fast-mode shock is not the only shock to be transmitted into the cloud. 
A slow-mode shock is trailing the fast-mode shock. As it moves much 
more slowly than the fast-mode shock, it remains close to the boundary of the 
cloud. Behind the slow-mode shock, the magnetic pressure decreases and there 
is nothing that prevents the gas from compressing as it cools. 
Figure~\ref{fig:pa_phase} clearly shows the rapid condensation due to the  
slow-mode shock. In a few Myr the gas behind the slow-mode shock cools to 
the thermally-stable cold phase. For the gas behind the fast-mode shock that 
is not processed by the slow-mode shock, this process occurs much more slowly. 

Thus, a dense, cold layer quickly forms at the cloud boundary with the highest
densities on the upstream parts of the cloud where the shock first hits the 
cloud (see Fig.~\ref{fig:paboundary}). While the cooling behind the slow-mode
shock is thermally unstable, the dense shell is subject to Kelvin-Helmholtz 
and Rayleigh-Taylor 
instabilities, even though the dynamical instabilities are mostly suppressed 
by the strong magnetic field \citep[e.g.][]{C61}. Hence, the shell 
breaks up into dense fragments. The densest clumps have number densities of 
$\approx 10^3$~ cm$^{-3}$ and are a few parsec in size. Note that we found 
roughly the same values for the higher resolution 2D simulation in Paper~I. 
For lower resolution 2D simulations the densities are not so high. However,
the clumps of the 2D simulations are essentially axisymmetric rings 
that break up into smaller, higher density clumps in the 3D simulations. 
These cold, dense clumps contain several hundreds of solar masses each and are 
gravitationally unstable. Thus, such clumps are likely precursors of 
massive stars.

\subsection{Perpendicular shock}
The dynamical evolution of cloud interacting with a perpendicular
shock is in many ways similar to that of one interacting with a parallel shock. 
As the intercloud shock sweeps around the cloud, a transmitted fast-mode shock 
propagates through the cloud making the cloud material thermally unstable. 
Figure.~\ref{fig:pe_phase} shows the distribution of mass in the $p-n$ phase 
space at two different times. The compression by a weak 
perpendicular shock is somewhat smaller than that by a parallel shock. 
The compression ratio changes from 2.70 for a parallel shock to 1.87
for a perpendicular shock. The gas behind the transmitted shock 
is therefore only marginally cooler and less dense. 
 
\begin{figure*}
\includegraphics[width = 8.4 cm]{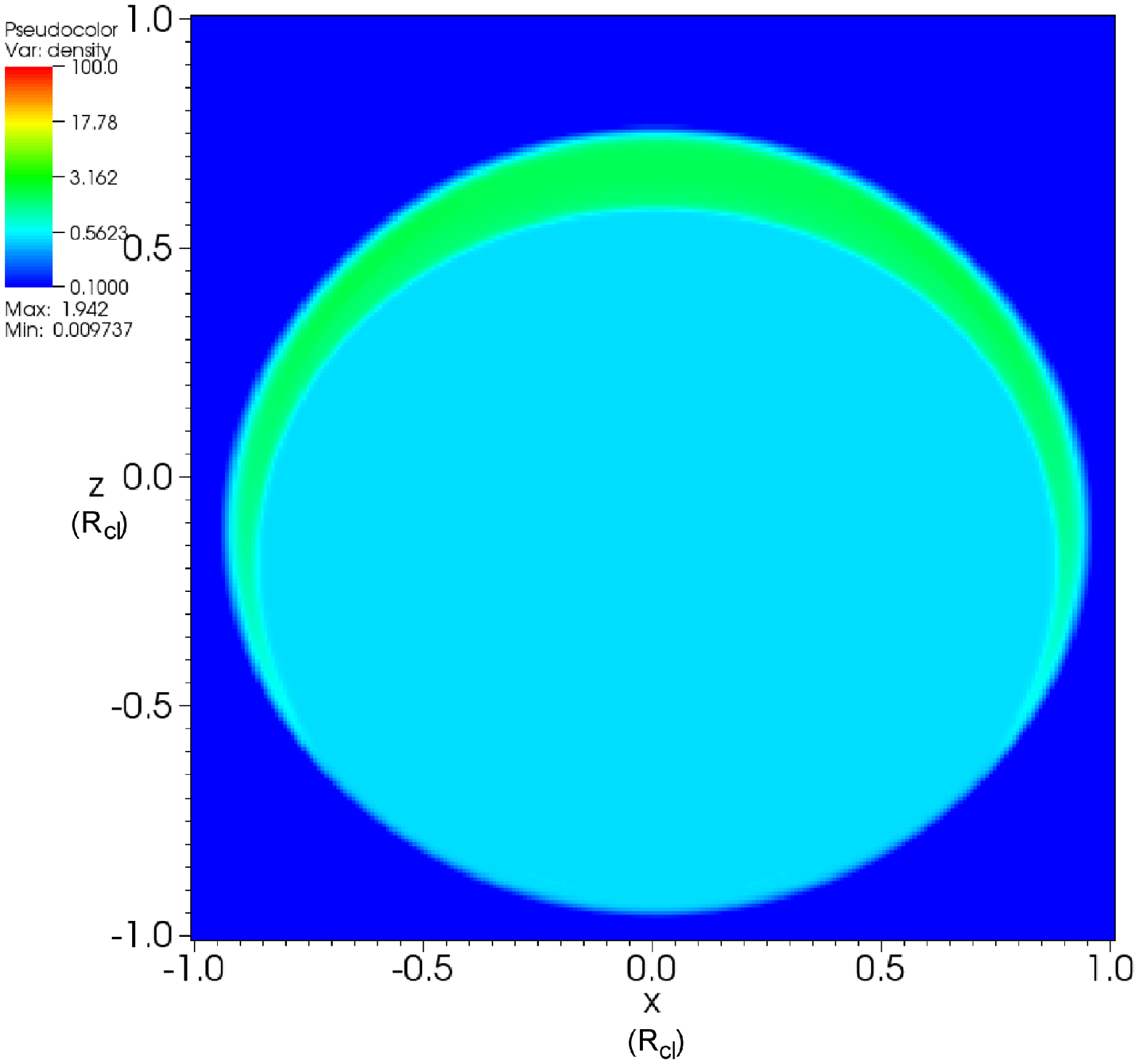}
\includegraphics[width = 8.4 cm]{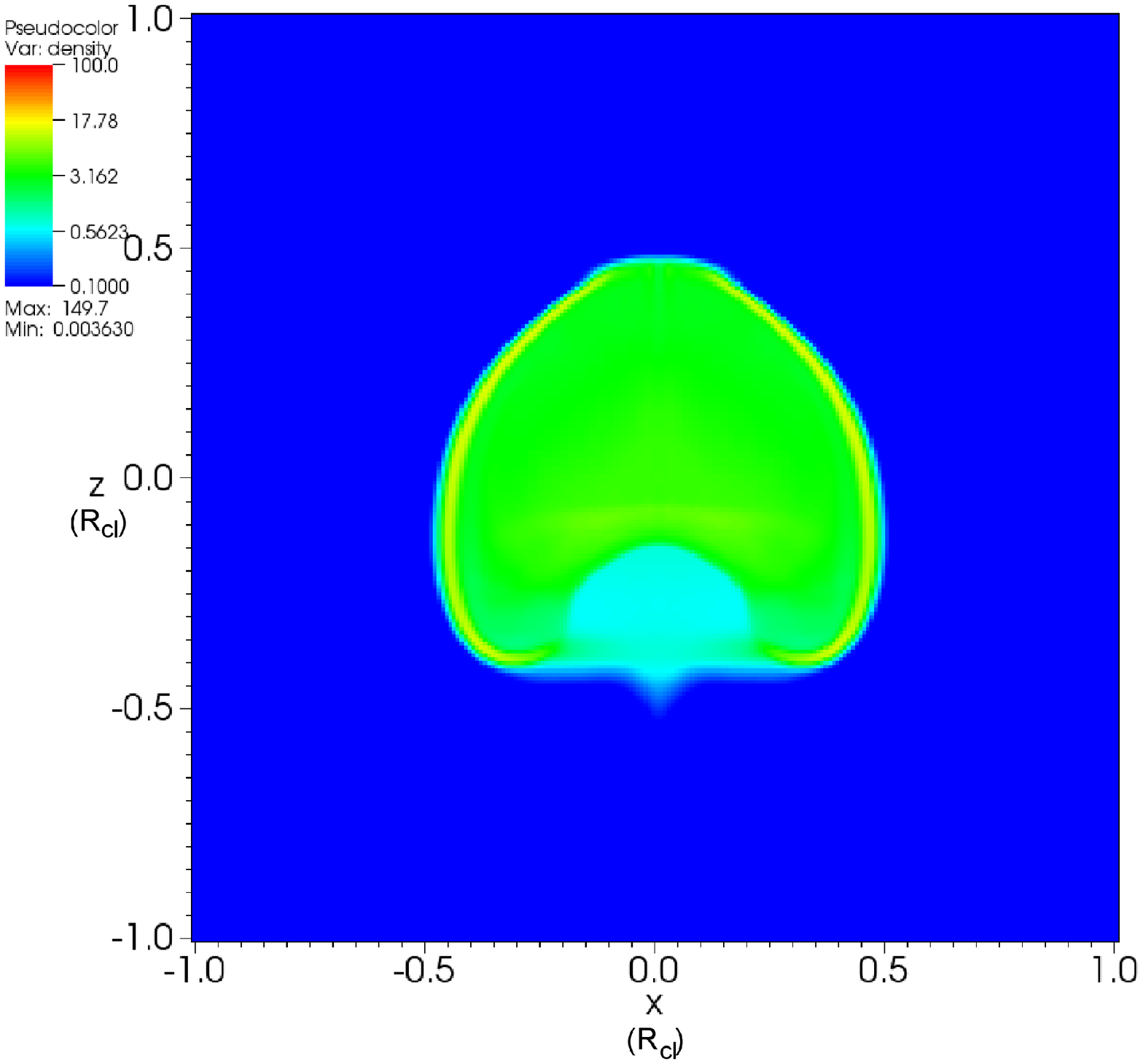}
\caption{
Similar to Fig.~\ref{fig:paboundary}, but for the perpendicular shock model.}
\label{fig:peboundary}
\end{figure*}

\begin{figure*}
\includegraphics[width = 8.4 cm]{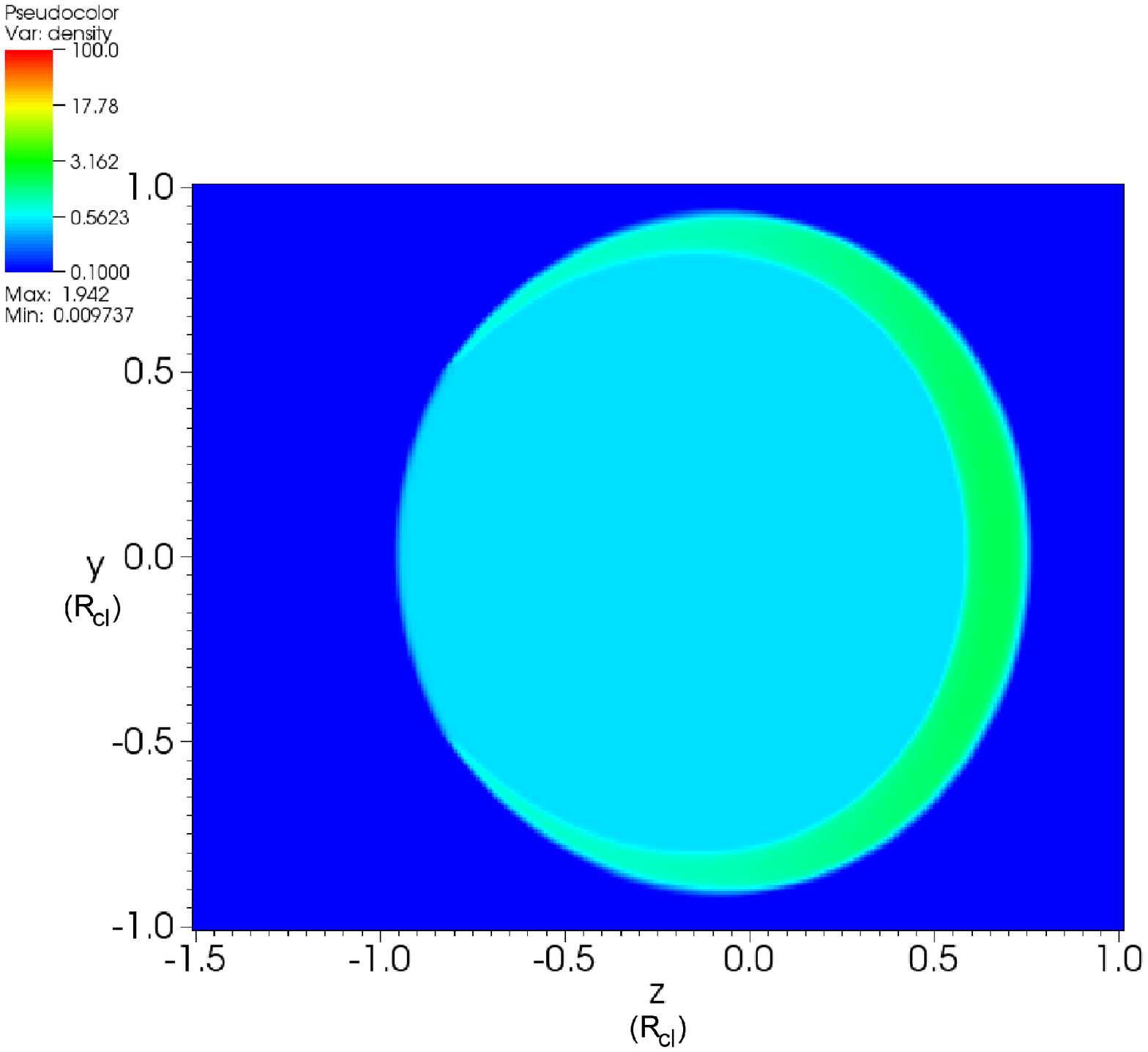}
\includegraphics[width = 8.4 cm]{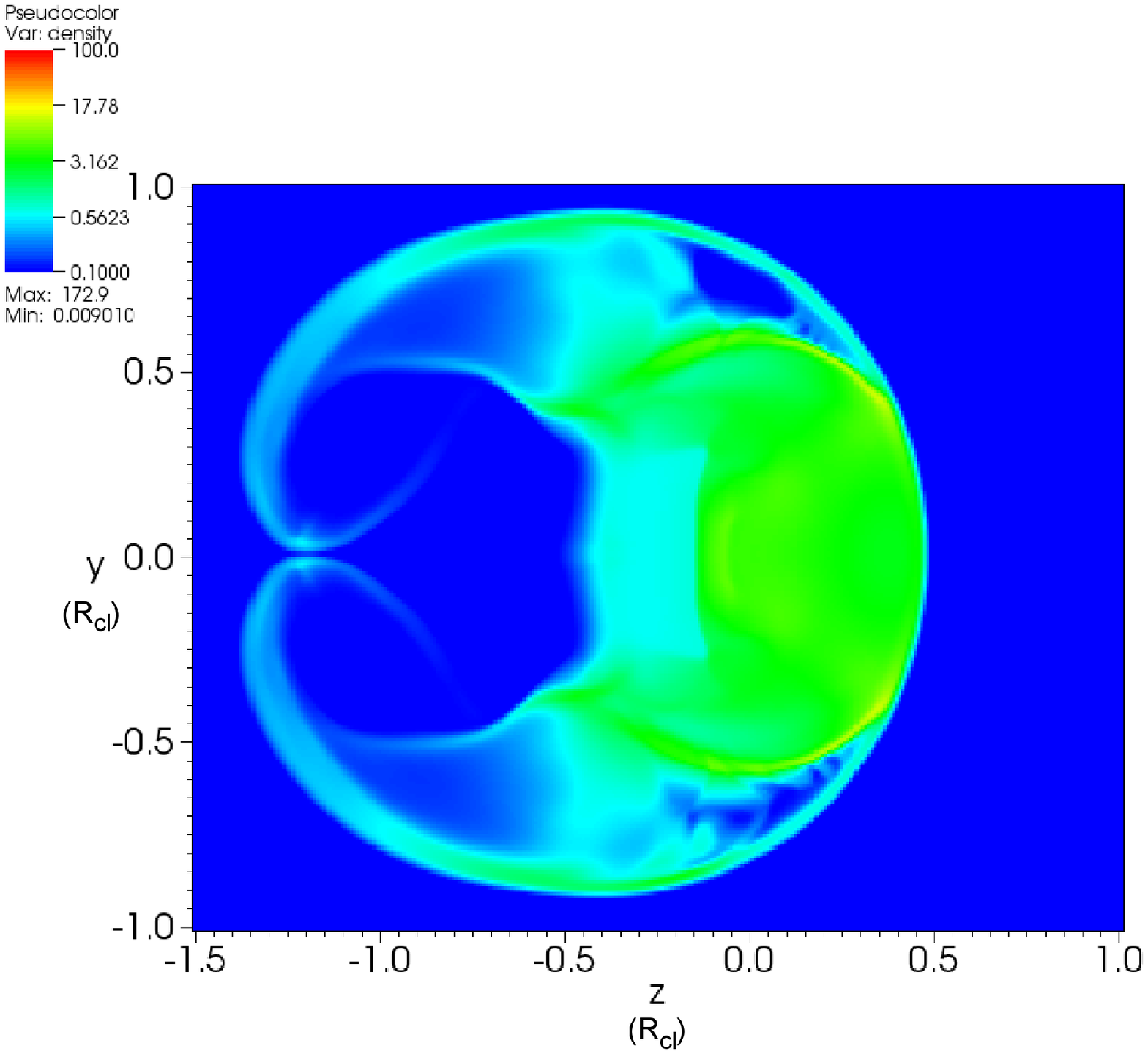}
\caption{
Similar to Fig.~\ref{fig:peboundary}, but the slice is now in the 
$x=0$ plane.}
\label{fig:peboundary_x}
\end{figure*}

However, contrary to what occurs in the parallel case, a slow-mode shock 
does not arise in the upstream parts of the cloud. This can be easily seen in 
Figs.~\ref{fig:peboundary} and \ref{fig:peboundary_x} which show the logarithm
of the density for the perpendicular case as a function of position
for different slices and different times. Slow-mode waves, and therefore 
slow-mode shocks, do not propagate perpendicular to the magnetic 
field. However, as the intercloud shock sweeps across the cloud, a 
slow-mode shock does arise at the sides of the cloud. This slow-mode 
shock is not as strong as in the parallel case. Thus, it does not trigger 
a condensation near the boundary. Figure~\ref{fig:pe_phase}
shows that the transition from warm, rarefied gas to cold, dense gas is rather
smooth with most of the gas near equilibrium. As discussed in 
Sect.~\ref{sect:parallel}, such a situation can initiate the formation of 
dense, cold clumps embedded in warm, rarefied gas \citep{KI06}.

The boundary layer in the perpendicular shock model is thus not as dense as 
in the parallel one. The maximum number density is an order of magnitude 
smaller $\approx 180$~cm$^{-3}$. Furthermore, the boundary layer is 
not subject to strong dynamical instabilities. While the boundary fragments 
into small, high-density clumps in the parallel model, it does not for 
the perpendicular model. Dynamical instabilities, notably the
Kelvin-Helmholtz instability, take longer to develop due to a lower velocity
shear and lower density, and are suppressed due to the 
strong magnetic field.

As seen in Figs.~\ref{fig:peboundary} and \ref{fig:peboundary_x}, which are 
of the logarithmic density for different slices, instabilities are 
generated after some time, but only in a specific 
plane, i.e. in the $x=0$ plane. It is the only direction for which a shear 
layer forms in the perpendicular model. In contrast, in the parallel shock
model the shear layer covers the entire cloud. The reason 
for this directional preference is found through the consideration of 
the vorticity equation, i.e. 
\[
 \frac{\partial{\bf \omega}}{\partial t} = \nabla \times ({\bf v} \times
 {\bf \omega}) + \frac{1}{\rho^2}
  \nabla \rho \times \nabla(p+\frac{B^2}{2}) - \frac{1}{\rho^2}
  \nabla \rho \times \left[({\bf B}.\nabla){\bf B}\right]
\]
where ${\bf \omega} = \nabla \times {\bf v}$ is the vorticity. While the first 
term on the right hand side describes the stretching of vorticity, 
the second is the baroclinic term and accounts for the changes in the 
vorticity due to the intersection of density and thermal and magnetic pressure 
surfaces. The third term relates to the changes due to the interaction of 
the density gradient and magnetic tension forces. For flows with $\beta \gg 1$,
the generation of vorticity is dominated by the thermal pressure contribution
to the baroclinic term. This term
does not introduce any directional preference. For the parallel shock model,
the post-shock flow has a large $\beta$ and vorticity is generated 
all around the cloud. Figure~\ref{fig:beta_dep} shows the downstream
plasma beta as a function of the angle between the shock normal and the 
magnetic field. For flows with 
$\beta \leq 1$, the contribution of the magnetic field to the vorticity 
generation cannot be neglected. Magnetic tension forces then annul the 
generation of vorticity by the baroclinic term. It is thus clear that 
in regions where the magnetic field lines are being bent, no shear layer 
arises. For the perpendicular shock model, this is exactly what happens.
The post-shock flow behind the intercloud shock drags the magnetic field 
with it along the $z$-direction. The field lines anchored in the cloud are 
bent and stretched in the $x,z$-plane near the cloud surface. Hence,
less vorticity is generated in the $x,z$-plane and a shear layer 
is only present in the $x=0$ plane.

\subsection{Oblique shock}
The perpendicular and parallel shock models describe the extreme cases of the 
shock-cloud interaction. Therefore, we can expect that the dynamical evolution 
for oblique shocks lies in between those of the perpendicular and parallel
models.

\begin{figure}
\includegraphics[width = 8.4 cm]{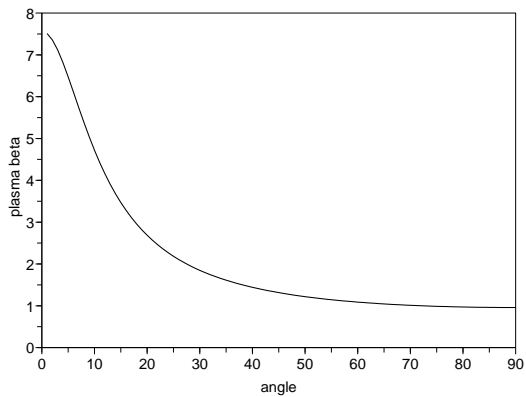}
\caption{Plasma beta of the gas downstream the intercloud shock as a function 
of the angle between the magnetic field and shock normal.}
\label{fig:beta_dep}
\end{figure}

For an angle of 45$^o$ between the shock normal and the magnetic field, 
the evolution of the cloud is similar to that of the perpendicular shock model. 
Although the transition of the gas from the warm to the cold phase is not as 
smooth as for the perpendicular shock model, it still progresses along the 
equilibrium curve. The maximum densities in the cloud 
are somewhat higher than for the perpendicular shock model. As for that model, 
the post-shock $\beta$ is $\approx 1$ (see Fig.~\ref{fig:beta_dep}). Thus,
magnetic tension again prevents the formation of a shear layer surrounding 
the cloud. The preferential direction of shear layer is perpendicular 
to the initial magnetic field and the flow direction.     

The 15$^o$ oblique shock model does not resemble the geometrical evolution 
of the perpendicular shock-cloud interaction. Instead, the evolution better
matches that of the parallel shock model as expected from the value of 
$\beta$ of the 
gas downstream the intercloud shock (Fig.~\ref{fig:beta_dep}).
As the intercloud shock sweeps around the cloud, a shear layer forms all around 
the cloud. However, a higher level of shear is found in the plane perpendicular 
to the magnetic field and the flow direction. As for the parallel shock
model, there is a rapid condensation behind the slow-mode shock and a dense
boundary layer forms. This dense boundary layer, however, does not fragment
due to instabilities as the small transverse component of 
the magnetic field has a stabilising effect. Although the geometrical 
features of the cloud resemble those of the parallel shock model, the cloud 
properties resemble those of the perpendicular model 
(see Sect.~\ref{sect:properties}). 

The shock-cloud interaction can thus be separated into two evolutionary tracks,
i.e. one that is quasi-parallel and one that is quasi-perpendicular. 
From Fig.~\ref{fig:beta_dep}, 
we see that the transition is associated with 
the value of $\beta$ downstream of the intercloud shock. For parallel shocks 
the postshock flow has a $\beta > 1$. Hence, the magnetic field is dynamically 
unimportant in the external gas which allows a vorticity layer to form around 
the cloud. For angles $> 20^o$, the magnetic field is important and 
suppresses the generation of vorticity and shear layers. 
   
\section{Discussion of cloud properties}\label{sect:properties}
\subsection{Size and density}
In order to study the properties of the different shock-cloud interaction 
models, we use diagnostics similar to those used by previous authors, 
\citep[e.g.][]{Metal94,SSS08}. Specifically, we use the 
density-weighted average of variables such as the plasma $\beta$ and the 
density using Eq.~\ref{eq:average}.

\begin{figure}
\includegraphics[width = 8.4 cm]{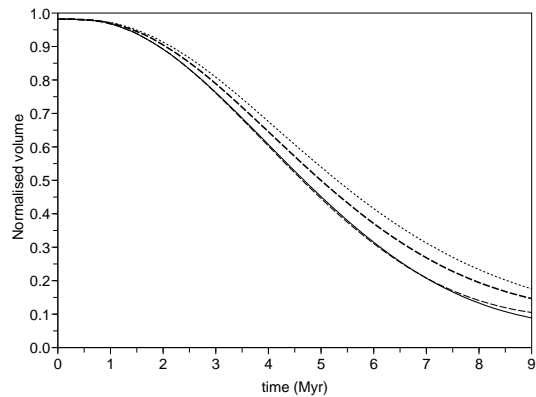}
\caption{Mean cloud volume derived from the mass-weighted moments 
along the different axes for the parallel (solid), perpendicular (dotted)
and oblique shock (thick dashed for 45$^o$ and dashed for 15$^o$) cases. 
The volume is normalised to the initial cloud size.}
\label{fig:volume}
\end{figure}

Furthermore, we define 
\[
        a = \left[5 \left(<x^2> - <x>^2 \right)\right]^{1/2}.
\]
along the $x$-axis, with similar expressions for analogous quantities defined
along the $y$ and $z$-axes. 
This gives us the axes for an ellipsoid with a similar mass distribution 
as the cloud's which are used to follow geometrical changes of the cloud. 
Figure~\ref{fig:volume} shows the temporal evolution of the volume 
of the cloud. The parallel shock compresses the cloud the most, while the 
perpendicular shock is the least compressive. At $t_{cc}$ the volume 
of the cloud is only a tenth of its original volume.
The ram pressure exerted 
by the post-shock flow of the perpendicular shock is smaller than that for 
the parallel shock, as both the post-shock density and gas velocity 
(in the lab frame) 
are lower. (A parallel shock compresses gas more than a perpendicular shock
with the same Mach number.) Furthermore, work needs to be done to compress 
the transverse magnetic field. Similarly, it is not surprising to find 
that the 45$^\circ$ oblique shock squeezes the cloud faster than the 
perpendicular shock, but slower than the parallel shock. For smaller 
angles we expect the confinement of the cloud to converge to the 
parallel shock model. Indeed, the volume of the cloud in the  
15$^\circ$ oblique shock model changes as for the parallel shock model 
with only a small deviation at later times.  

Figure~\ref{fig:volume} seems to show a minimum in the volume 
near 9~Myr. From the 2D axisymmetric simulations, we know that a minimum arises 
as the fast-mode shock is reflected at the centre of the cloud at 
around $t_{cc} (\approx$ 8.3 Myr) and propagates back to the boundary.  
This shock reflection initiates a re-expansion phase in the direction 
perpendicular to the post-shock flow direction \citep[see][Paper~I]{Metal94}
Along the flow direction, the compression of the cloud continues and 
the cloud ends up as a thin disc after 1.5-2 $t_{cc}$. This means that 
the geometrical shape of the cloud resembles more closely an oblate 
spheroid than 
a sphere. The ellipticity of the cloud changes from 0 initially (i.e. a sphere)
to roughly 0.5 at $t_{cc}$. As mentioned earlier,
we do not follow this expansion phase and stop the simulation  shortly 
after the cloud-crushing time $t_{cc}$.  

\begin{figure}
\includegraphics[width = 8.4 cm]{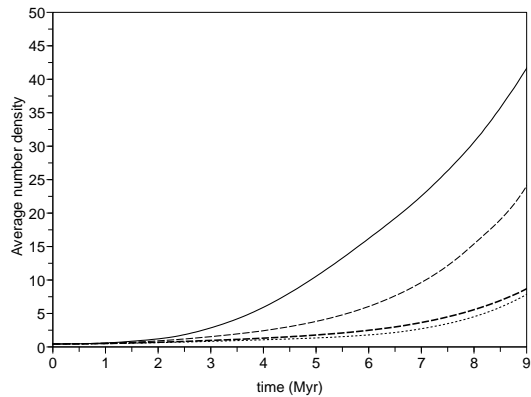}
\caption{rms number density of the cloud for the parallel (solid), 
perpendicular (dotted) and oblique shock (thick dashed for 45$^o$ 
and thin dashed for 15$^o$) cases.}
\label{fig:rhoav}
\end{figure}

Figure~\ref{fig:rhoav} gives the evolution of the root-mean-square (rms)
number density.  Although the volumes of the cloud are only different by 
a factor of 2 as can be seen from Fig.~\ref{fig:volume}, the 
rms density changes more significantly. 
At $t_{cc}$, the rms number density for the parallel model is $\approx
40$~cm$^{-3}$, while it is only 10~cm$^{-3}$ for the perpendicular model.
This large variation is due to the large difference in the density of the 
boundary layer between the models (see Sect.~\ref{sect:results}). 
Only models with small angles between the shock normal and the 
magnetic field have mean densities similar to those of observed GMCs. 
Large angle models only reach mean densities similar to diffuse HI clouds.
Such a dependence of the density structure on the magnetic field orientation 
is also observed by \citet{HSH09}. 

\begin{figure}
\includegraphics[width = 8.4 cm]{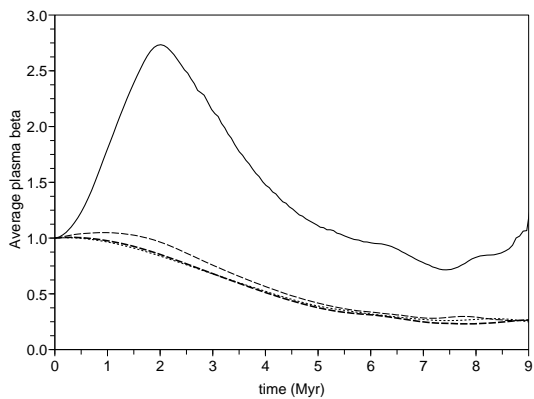}
\caption{Mass-weighted plasma beta of the cloud for the parallel (solid), 
perpendicular (dotted) and oblique shock model (thick dashed for 
45$^o$ and thin dashed for 15$^o$). }
\label{fig:betaav}
\end{figure}

\begin{figure}
\includegraphics[width = 8.4 cm]{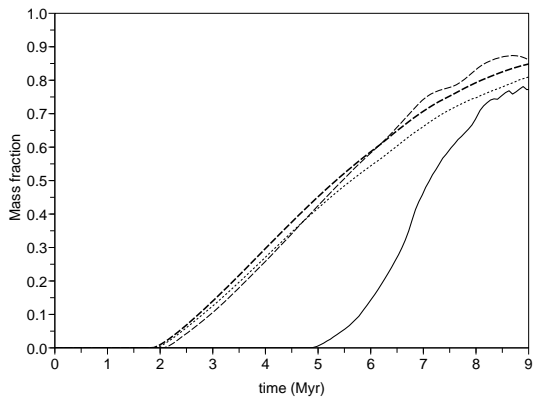}
\caption{Mass fraction of the cloud with $\beta < 0.1$ for 
the parallel (solid), perpendicular (dotted)
and oblique shock model (thick dashed for 45$^o$ and thin dashed for 15$^o$).}
\label{fig:massbelow}
\end{figure}

\subsection{Thermal and magnetic pressure}
Observations show that molecular clouds are magnetically dominated with 
plasma $\beta$ of the order 0.04 - 0.6 \citep{CHT03}. Figure~\ref{fig:betaav}
shows the mass-weighed mean of $\beta$. This weighted mean value of 
$\beta$ for the parallel shock model does not lie within 
the observed range. During the early stages of the 
evolution the gas behind the slow-mode shock has a large thermal pressure, 
while the magnetic pressure is small. Hence, the plasma $\beta$ is large for 
the high-density gas in the boundary layer (which dominates the mean value 
of $\beta$), i.e. $\beta \approx 5$. After about 2.5~Myr, the weighted 
mean value of $\beta$ decreases as the thermally unstable gas behind the 
fast-mode shock becomes magnetically dominated. Although the weighted mean 
value of $\beta$ is now around unity, Fig.~\ref{fig:massbelow} shows that a 
significant mass fraction of the cloud has $\beta < 0.1$.  At $t_{cc}$, 
about 75\% of the gas is significantly magnetically dominated.  

The other models do produce a cloud with a weighted mean $\beta$ of the order
of 0.4 (see Fig.~\ref{fig:betaav}). The main reason for this has to do 
with the role played by the transverse component of the magnetic field. While 
the thermal pressures behind
the fast-mode and slow-mode shock are not as high as in the parallel shock
model (i.e. the transition from the warm phase to the cold phase is smoother),
the magnetic field is significantly compressed behind the fast-mode shock.
The combined result of these effects produces a lower $\beta$ in both the 
boundary layer and within the cloud. Figure~\ref{fig:massbelow} indeed shows
that magnetically-dominated gas appears much earlier for shock models with
a transverse component to the magnetic field.  After only 3~Myr, 
already 10\% of the total mass in these models is magnetically dominated.   

Our simulations show that, to produce clouds with magnetically-dominated 
high-density gas, the angle between the shock normal and the magnetic field
must be small. While perpendicular shocks produce magnetically-dominated gas 
at low densities, high-density clumps with $\beta \gg 1$ arise in the 
parallel shock models.

\subsection{Fragmentation}
Our models show that large fractions of the cloud are magnetically dominated.
This provides the ideal conditions for MHD waves to generate high-density 
clumps and cores within the cloud \citep{FH02,VFH06,VFH08}. 
For shock models with 
a transverse component of the magnetic field, this process initiates earlier
suggesting a higher degree of fragmentation. On top of that, another process 
is effective in these models. As the transition from the thermally-stable warm 
phase to the cold one follows the unstable part of the equilibrium 
curve, small perturbations can initiate the formation of dense, cold clumps 
embedded in warm, diffuse gas \citep{KI06}. Unfortunately, we cannot follow
these clump and core formation processes as our resolution is insufficient.

While a low resolution is partly to blame for the low amount of fragmentation, 
the uniform initial conditions of the cloud also play an important role. In 
the colliding flow-driven models of \citet{Hetal08} and \citet{HSH09}, 
the generation of cold dense cores and clumps relies on seeded perturbations 
in either the incoming flow or at the collision front. 
Without these perturbations, the collision 
region remains roughly uniform and fragmentation occurs on very long 
timescales. Therefore, it can be expected that the introduction of 
perturbations within the cloud and at the edge of the cloud would 
produce much more fragmentation.

Furthermore, our simulations do not include the effect of self-gravity.
While self-gravity is dynamically unimportant for the global evolution of the 
cloud, i.e. the mass of the cloud is much smaller than its Jeans' mass, its 
effect will become important locally. For example, the high-mass boundary layer
clumps of the parallel shock model have masses that exceed their Jeans mass. 
We will investigate the effect of self-gravity on the cloud evolution in 
a later paper.

\begin{figure}
\includegraphics[width = 8.4 cm]{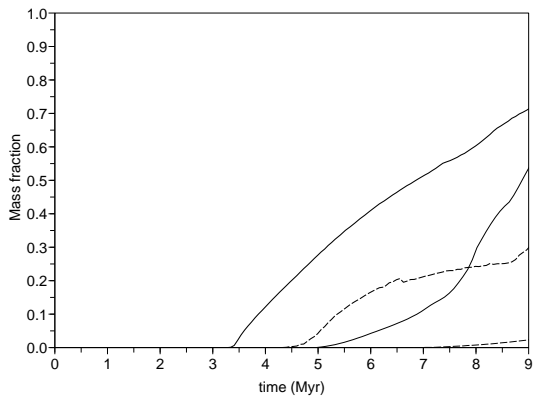}
\caption{Mass fraction of cold (solid) and molecular (dashed) gas 
within the cloud for the parallel (thick) and perpendicular (thin) shock model. 
Molecular gas is assumed to be present when the thermal gas pressure is 
higher than 2500$k$ and the gas temperature lower than 100K.}
\label{fig:molecular}
\end{figure}

\subsection{Molecular clouds}
Our simulations only describe the dynamical evolution 
of a cloud from warm atomic gas to cold atomic gas. 
Figure~\ref{fig:molecular} shows the temporal evolution of the 
cold gas mass fraction for the parallel and perpendicular shock model. 
Cold gas arises earlier in the parallel shock model than in the 
perpendicular one, but after 9 Myr more than 50\% of the initial cloud
mass is in the thermally stable cold gas phase. Although we have not included
a description of molecular cooling, nor do we follow the cloud chemistry, 
we can roughly estimate how much gas is converted from the cold atomic gas 
to molecular gas. \citet{WBS95} find that the average number density of
H$_2$ in the CO clumps of the Rosette Molecular Cloud is 
$\approx 220~{\rm cm^{-3}}$. With typical excitation temperatures between
10 and 20 K, the thermal gas pressure of these clumps is roughly 2500k.
Therefore, we assume that any gas parcel in the shocked cloud with a thermal 
pressure higher than 2500k and a temperature below 100K will become 
molecular given enough time (see below). Using this criterion, we find that
the parallel shock model generates a molecular cloud as half of the cold gas
becomes molecular (see Fig.~\ref{fig:molecular}). In the perpendicular cloud 
model only a small fraction of the cold atomic gas is converted into molecular
gas. This suggests that the perpendicular shock model only produces 
a diffuse HI cloud.

\begin{figure*}
\includegraphics[width = 8.4 cm]{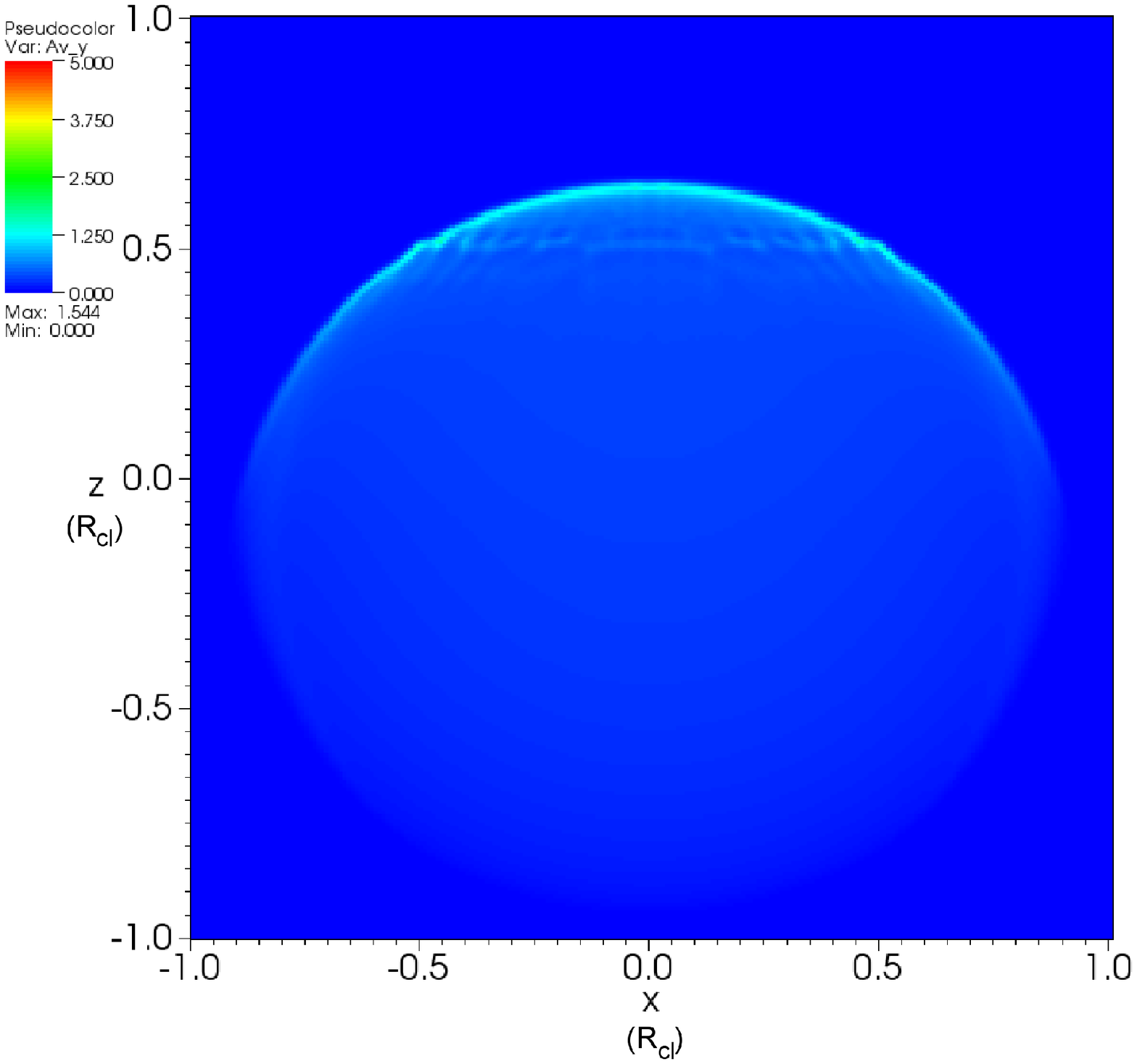}
\includegraphics[width = 8.4 cm]{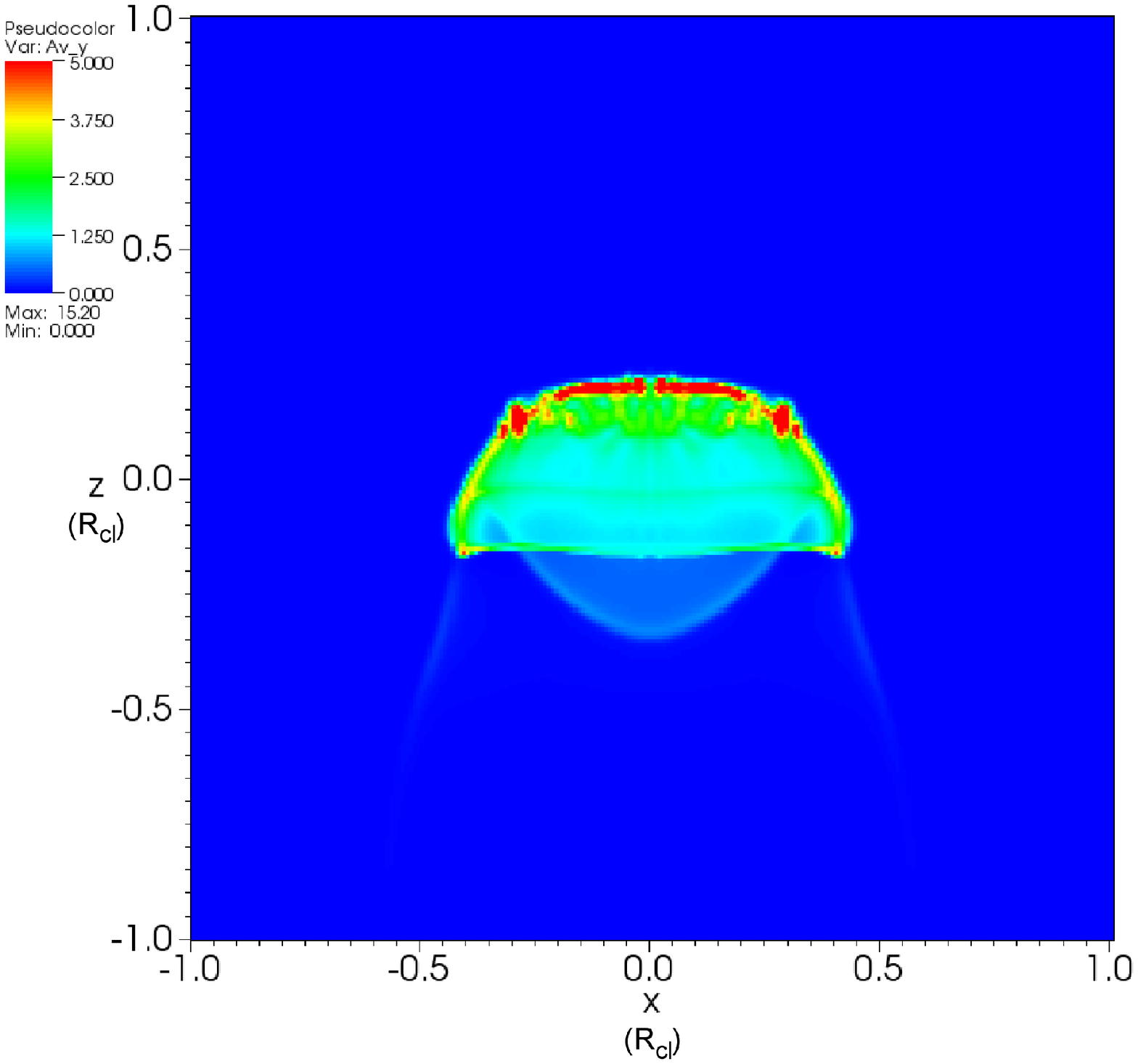}
\caption{Visual extinction along the $y$-axis for a 
parallel shock interacting with a diffuse cloud. The left panel 
show the visual extinction  at 2.5 Myr and the right panel at 9 Myr.}
\label{fig:Av_pa}
\end{figure*}

The above result is only valid if the time scale for the formation of 
molecules is short and if the physical parameters of the formation process 
are met.
\citet{Getal09} use high-resolution 3D simulations of turbulent interstellar 
gas to follow the formation and destruction of molecular hydrogen and CO. 
They find that most CO forms within 2-3~Myr for dense, turbulent gas, 
while the formation of H$_2$ is even faster, i.e. within 1-2~Myr. Their results 
indicate that once large enough spatial and column densities are reached, 
the conversion from atomic to molecular gas is rapid. A good indictor for 
the formation of molecules is the visual extinction $A_V$, 
which can be expressed as \citep[e.g.][]{CC04}
\[
     A_V = \frac{N_{H}}{1.80 \times 10^{21} {\rm cm^{-2}}}.
\]
For regions with $A_V \gtrsim 0.5$ and high local densities, we can 
then expect that molecules are present.  Figure~\ref{fig:Av_pa} shows that 
the visual extinction is 
already high early on in the parallel shock model. These regions also 
correspond to high-density regions, as can be seen in Fig.~\ref{fig:paboundary}.
This model thus most likely produces a molecular cloud. Also,
note the similarity of the column density plot for the parallel shock model
with the emission map of the W3 GMC (see Fig.~6 of Paper~I). A more structured 
inner cloud can be expected with a non-uniform initial condition and a 
higher resolution. 

While the 15$^\circ$ oblique shock model also produces high column densities
which coincide with high density regions, the perpendicular and the 45$^\circ$
models do not. Models with large transverse components of the magnetic field 
produce diffuse HI clouds instead of molecular clouds. However,
this conclusion only holds for our current simulations.  A higher 
resolution and inclusion of small-scale perturbations potentially would
produce higher density clumps for these models. 

\section{Summary}\label{sect:conclusions}
In this paper we have presented 3D simulations of the interaction of a 
weak, radiative shock with a magnetised, diffuse atomic cloud. 
The interaction of the shock induces the transition of the cloud from
the thermally warm atomic phase to the cold one. By
modelling the shock-cloud interaction in 3D, we are able to study
the effect of different magnetic field orientations including parallel, oblique
and perpendicular to the shock normal.
 
Contrary to the strong, adiabatic shock models of \cite{SSS08}, we 
find that the structure of the shocked cloud differs significantly with 
the magnetic field orientation. The shock-cloud interaction can be
separated into two distinct classes, i.e. a quasi-parallel one and 
a quasi-perpendicular one. In the quasi-parallel shock models 
high-density clumps are generated in the boundary layer surrounding
the cloud. As the visual extinction of the gas is also high, the resulting 
cloud is most likely molecular. The quasi-perpendicular shock models,
however, only produce low-density clouds resembling HI clouds. This result is
similar to the one of \cite{HSH09}.
 
All our models show that the shocked cloud becomes magnetically dominated 
after a few Myr. Although this provides the ideal conditions for the 
formation of dense clumps and cores \citep{FH02, VFH06, VFH08}, we do not 
see this happening in our simulations. This can be partly ascribed due to 
the assumption of an initially quiescent, uniform and spherical cloud. 
From the colliding flow-driven models we know that the inclusion of 
small-scale perturbations will generate a higher degree of structure. 
Increasing the resolution can have a similar effect \citep{YFC09}. 
In a subsequent paper we will investigate the effect of small-scale 
perturbations and a higher resolution on the formation of clumps and 
cores. 

\section*{Acknowledgements}
We  thank the anonymous referee for a report that helped to
improve the manuscript. The authors gratefully acknowledges STFC for the
financial support.

\label{lastpage}

\end{document}